\documentclass[aps,prl,twocolumn,10pt,showpacs,superscriptaddress,amsmath,amssymb,nobibnotes]{revtex4-1}

\usepackage{graphicx}
\usepackage{amsfonts}    % need for subequations
\usepackage{graphicx}   % need for figures
\usepackage{verbatim}   % useful for program listings
\usepackage{color}      % use if color is used in text
\usepackage{subfigure}  % use for side-by-side figures
\usepackage{hyperref}   % use for hypertext links, including those to external documents and URLs
\usepackage{natbib}
\usepackage{booktabs}
\usepackage{threeparttable}
\usepackage{dcolumn}
\usepackage{multirow}
\begin{document}
\title{Experimentally undoing an unknown single-qubit unitary}
\author{Qin Feng}
\altaffiliation{These authors contributed equally}
\affiliation{State Key Laboratory of Optoelectronic Materials and Technologies and School of Physics, Sun Yat-sen University, Guangzhou 510006, China}

\author{Tianfeng Feng}
\altaffiliation{These authors contributed equally}
\affiliation{State Key Laboratory of Optoelectronic Materials and Technologies and School of Physics, Sun Yat-sen University, Guangzhou 510006, China}

\author{Yuling Tian}
\affiliation{State Key Laboratory of Optoelectronic Materials and Technologies and School of Physics, Sun Yat-sen University, Guangzhou 510006, China}

\author{Maolin Luo}
\affiliation{State Key Laboratory of Optoelectronic Materials and Technologies and School of Physics, Sun Yat-sen University, Guangzhou 510006, China}

\author{Xiaoqi Zhou}
\email{zhouxq8@mail.sysu.edu.cn}
\affiliation{State Key Laboratory of Optoelectronic Materials and Technologies and School of Physics, Sun Yat-sen University, Guangzhou 510006, China}
\date{\today}
%\pacs{03.67.Lx, 03.67.Pp, 03.65.Vf}

%\begin{spacing}{1.2}

\begin{abstract}
 
% Inverse unitary operation, an important operation in quantum dynamics and quantum information processing, can cancel the unexpected influence of unitary evolution in quantum systems. Given a  unitary operation without any specific description, however, it is a hard and challenging task to realize the inverse operation of the unitary. Recently, a universal quantum circuit was proposed in [Phys.Rev.Lett. 123, 210502 (2019)] to transform an arbitrary unknown but $d$-dimensional unitary $U$ into its inverse operation $U^{-1}$ with probability. In this letter, we report an experimental demonstration of reversing three unknown single-qubit unitaries $(d=2)$ by using linear optical system. The experimental results show the average fidelity of inverse unitaries is $F=0.9767\pm0.0048$ which is in good agreement with the theoretical prediction. 

Undoing a unitary operation, $i.e$. reversing its action, is the task of canceling the effects of a unitary evolution on a quantum system, and it may be easily achieved when the unitary is known. Given a unitary operation without any specific description, however, it is a hard and challenging task to realize the inverse operation. Recently, a universal quantum circuit has been proposed [Phys.Rev.Lett. 123, 210502 (2019)] to undo an arbitrary unknown $d$-dimensional unitary $U$ by implementing its inverse with a certain probability. In this letter, we report the experimental reversing of three single-qubit unitaries $(d = 2)$ by linear optical elements. The experimental results prove the feasibility of the reversing scheme,  showing that the average fidelity of inverse unitaries is $F=0.9767\pm0.0048$, in close agreement with the theoretical prediction.

% 只能做单比特 怎么说意义

\end{abstract}

\maketitle
%\section{Introduction}

In quantum theory, quantum operations are mathematical maps evolving quantum states in the Hilbert space. They play a crucial role in quantum computation and quantum information processing\cite{NC00}. Among quantum operations,  unitaries are those inherently reversible.  To recover or reset the original quantum state after evolution \cite{M18,Reset2} , or to cancel the action of a quantum circuit, one needs to undo a unitary operation, $i.e$. to reverse the original unitary. The inverse unitary operation, defined as the inverse physical process of the original unitary, can be calculated mathematically if the unitary is known. Indeed, given a known unitary operation $U$, one may easily find the inverse unitary $U^{-1}$ by a classical computer, and then decompose it into the tensor product of a series of fixed causal order and low-dimensional unitary quantum gates\cite{NC00}.  

Here we address a somehow different problem: can we implement the inverse operation $U^{-1}$ without having any knowledge about the unitary $U$, which is given as a black box? This task is in general very challenging, since the only direct approach is that of reconstructing the unknown unitary by quantum process tomography (QPT)\cite{CN97}, which however requires repeated preparations of the system and it is characterized by low efficiency.

Recently, attention has been paid to reversing methods based on gate estimation. For instance, Chiribella and Ebler\cite{CE06} developed a semidefinite programming approach to optimize quantum networks, and applied it to engineer inverse gates. Moreover, they calculated the maximum expected fidelity in obtaining the inverse operation of an arbitrary $d$-dimension unitary $U_d$ as $F=\frac{2}{d^{2}}$. With the purpose to correct the undesirable effects of a perturbation in an isolated quantum system, Sardharwalla \emph{et al}\cite{SC12} proposed a new inverse-free version of the Solovay-Kitaev theorem\cite{KSV02,DN05} which approximates inverse gates. However, estimation precision is limited and therefore the above schemes are only approximate and cannot realize exact inversion of unitary operations.

Quite recently, Quintino \emph{et al}\cite{QD19,QD019} proposed a probabilistic scheme to implement inverse unitary operation by gate teleportation scheme\cite{BB93,GC99}, thus providing a universal quantum circuit to exactly implement the inverse operation $U_d^{-1}$ of an arbitrary unknown $d$-dimensional unitary operation $U_d$. The scheme can be divided into two parts. The first part consist in realizing the complex conjugate of an arbitrary $d$-dimensional unitary operation\cite{MS19}, whereas the second part involves the implementation of the transpose of the remain quantum operation. The probability of failure of this scheme decreases exponentially as the number of $U$ queries increases.

In this letter, we report a proof-of-principle experiment, implementing the inverse operation of a single-qubit unitary by linear optical elements. Our scheme exploits a pair of entangled photons as a quantum resource, carrying information about the target unitary to realize the inverse operation on the input qubit. The paper is organized as follows. At first, we introduce the scheme to realize the inverse operation of arbitrary single qubit unitary. Then, we illustrate our experimental implementation by linear optics, and present our results. Finally, we close the letter with some concluding remarks.

\begin{figure}[tb]
   \includegraphics[scale=0.6]{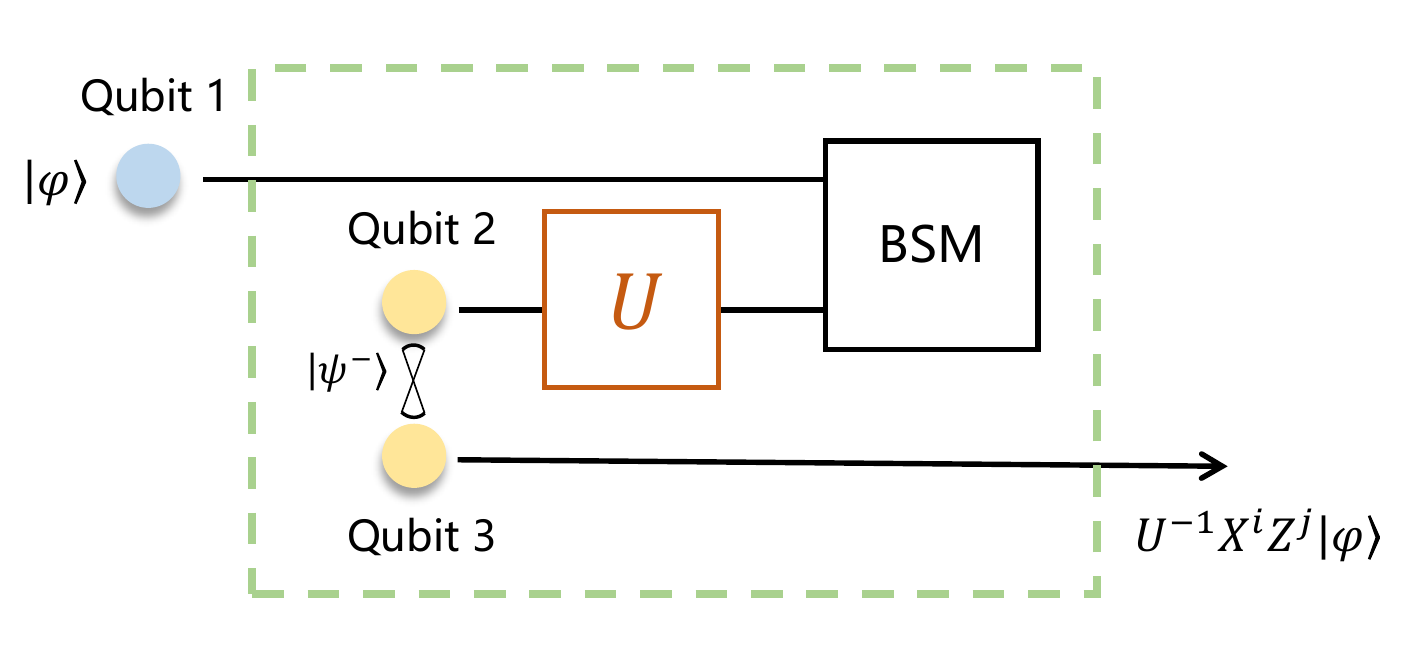}
   \caption{
   Quantum circuit to implement the inverse operation $U^{-1}$. The unitary $U$  is applied on qubit 2 of a singlet state $|\psi^- \rangle_{23}=\frac{1}{\sqrt{2}} (|0\rangle_2|1\rangle_3 -|1\rangle_2|0\rangle_3)$, and a Bell state measurement is performed on qubit 1 and qubit 2. The remaining qubit 3 is left in the state $ U^{-1}X^{i}Z^{j}|\varphi \rangle$, which leads to the desired gate operation $U^{-1}$ when $i=j=0$. 
   }
\end{figure}

\emph{Inversion of single-qubit unitary based on gate teleportation-} Here, we briefly review the general scheme for undoing an unknown single-qubit unitary. As illustrated in Fig. 1, given an arbitrary single qubit state $|\varphi \rangle$~(qubit 1), the inverse operation of the single-qubit unitary $U$ may be obtained on the final state of qubit 3 by performing $U$ on qubit 2 of the entangled resource. To be more specific, let us consider a singlet state $|\psi^- \rangle_{23}=\frac{1}{\sqrt{2}} (|0\rangle_2|1\rangle_3 -|1\rangle_2|0\rangle_3)$ as the quantum resource for teleportation. If the unitary $U$ is operated on qubit 2 of $|\psi^- \rangle_{23}$, we obtain the two-qubit entangled state
\begin{equation}
|\psi_U \rangle_{23} = (U\otimes I) |\psi^- \rangle_{23}.
\end{equation}
Since $U^{-1}U=I$,  eq.(l) can be rewritten as
\[
|\psi_U \rangle_{23} = (U\otimes U^{-1}U) |\psi^- \rangle_{23} =(I\otimes U^{-1})(U\otimes U) |\psi^- \rangle_{23}.
\]
It is worth noticing that when the same single-qubit unitary $U$ is performed on both qubits of $|\psi^- \rangle$, the system is left in the same quantum state $|\psi^- \rangle$, apart from a global phase which has been ignored. Hence, the two-qubit entangled state can be written as
\begin{equation}
|\psi_U \rangle_{23}= (I\otimes U^{-1})|\psi^- \rangle_{23}.
\end{equation}
\begin{figure*}[htbp]
  \center
   \includegraphics[scale=0.495]{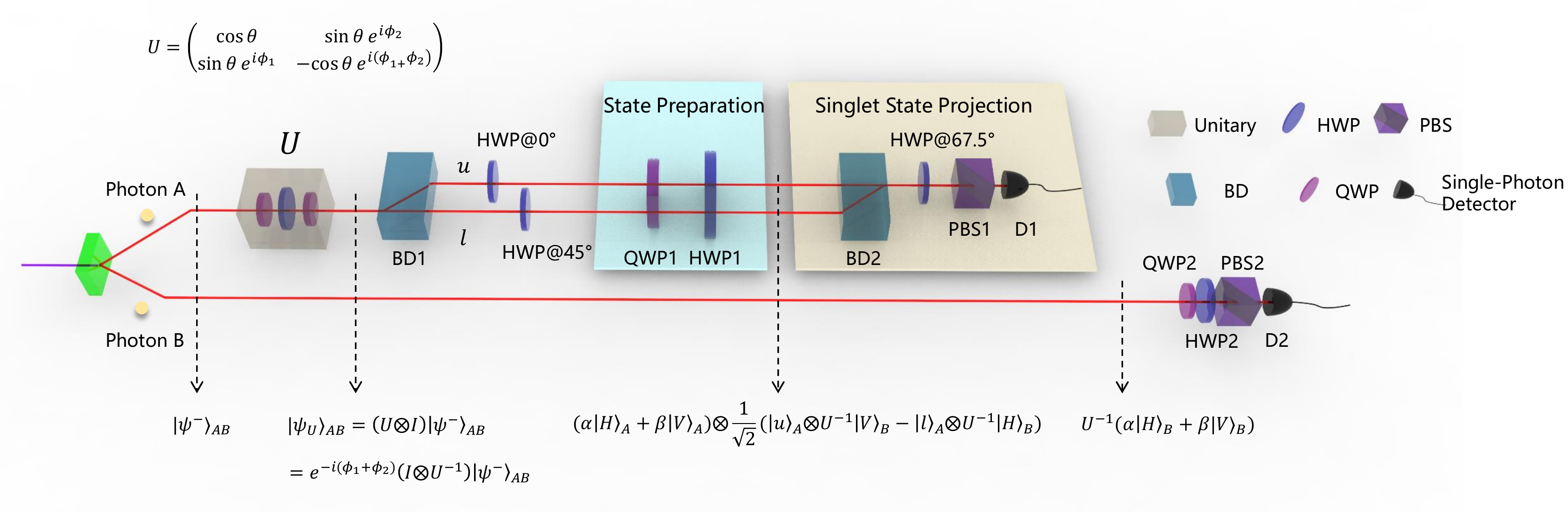}
   \caption{
   Experimental setup: The barium borate (BBO) crystal is pumped by a 405nm laser to generate a pair of polarization-entangled photons via type-II spontaneous parametric down-conversion (SPDC). The box, denoting an unknown unitary operation, is realized by two quarter waveplates (QWPs) and one half waveplate (HWP) on photon A. Photon A then passes through the beam displacer BD1, with the two HWPs set at $0^\circ$ and $45^\circ$ in upper path ($u$) and lower path ($l$) respectively, to achieve the conversion of degree of freedom between polarization and path of photon A. Therefore, the quantum state of the two photons can be written as $|H\rangle_A \otimes\frac{1}{\sqrt{2}}(|u\rangle_A\otimes U^{-1}|V\rangle_B-|l\rangle_A\otimes U^{-1}|H\rangle_B)$ . The polarization qubit of photon A, is then prepared in $|\varphi \rangle_{A} = \alpha |H\rangle_A + \beta |V\rangle_A$  by  QWP1 and HWP1. The polarization qubit and the path qubit of photon A are projected into the singlet state $\frac{1}{\sqrt{2}}(|H \rangle_{A}|l\rangle_A-|V \rangle_{A}|u\rangle_A)$ by BD2 with a HWP at $67.5^\circ$, polarization beam splitter PBS1 and D1. The remaining qubit 3, encoded in the polarization of photon B, is measured by a polarization analyzer consisting of QWP2, HWP2, PBS2 and D2.
   }
\end{figure*}
Eq.(l) and eq.(2) are equivalent. Thus, for the singlet state $|\psi^- \rangle$ performing the unitary operation $U$ on one qubit is equivalent to perform the inverse operation $U^{-1}$ on the other qubit. In order to achieve the inverse unitary operation on $|\varphi \rangle$,  a Bell state measurement (BSM) is required on qubit 1 and qubit 2, leaving qubit 3 in the state $|\varphi_{out} \rangle=U^{-1}X^{i}Z^{j}|\varphi \rangle$, where $X$ and $Z$ denote Pauli matrices and $i,j\in\{0,1\}$ are the outcome of the BSM. 
%Overall, we have that the quantum circuit of Fig.1(b) is equivalent to the quantum circuit of Fig.1(a). The area in the blue frame is the quantum teleportation circuit. As the BSM is performed on the input qubit and qubit 1, the information carried by the input state is teleported to qubit 2. Upon implementing the inverse operation, the final state of qubit 2 is given by eq.(4).
 Notice that given an arbitrary unitary $U$, its inverse $U^{-1}$  generally does not commute with and the Pauli operators $X$ and $Z$, and thus the circuit is inherently probabilistic.

The success probability of the scheme is $\frac{1}{4}$ with a single query of $U$. If qubit 1 and qubit 2  are projected onto $|\psi^- \rangle_{12}=\frac{1}{\sqrt{2}} (|0\rangle_{1}|1\rangle_{2} -|1\rangle_{1}|0\rangle_{2}) $, corresponding to the outcome $i=j=0$ in the BSM, the single-qubit inverse unitary operation is exactly performed on the output qubit 3, $ i.e.$   $|\varphi_{out} \rangle=U^{-1}|\varphi \rangle$.  On the other hand, if the outcome of BSM is not $i=j=0$, more queries are required, $ i.e.$ we must make use of $U$ again to apply the operation $Z^{-j}X^{-i}U$ on $|\varphi_{out} \rangle$ to recover $|\varphi \rangle$ and then restart the protocol again. The times needed to repeat the queries of $U$ increases approximately linearly, whereas the success probability increases exponentially. Upon registering the 00 outcome, the inverse unitary operation is exactly performed on the input state with unit fidelity.

\emph{Experimental setup and results-} Based on the theoretical scheme above, we now present our experimental linear optics implementation. As shown in Fig.2, a polarization-entangled singlet state of two photons $|\psi^- \rangle_{AB}=\frac{1}{\sqrt{2}} (|H\rangle_A|V\rangle_B-|V\rangle_A|H\rangle_B)$  degenerated at 810nm, is produced by pumping a type-II BBO crystal with an ultraviolet laser $@$405nm, where H~(V) denotes horizontal~(vertical) polarization. Photon A then goes through an unknown single-qubit unitary $U$, which can always be described as
\[
U=\left(
      \begin{array}{cc}
       cos\theta & sin\theta e^{i\phi_2}\\
       sin\theta e^{i\phi_1} & -cos\theta e^{i(\phi_1+\phi_2)}\\
      \end{array}
      \right).
\]
In our experiment, $U$ is realized by a sandwich-like structure composed of two quarter waveplates~(QWPs) and a half waveplate~(HWP). After photon A passing through $U$, the state of the two photons becomes 
\[
\begin{split}
|\psi_U \rangle_{AB} =& (U\otimes I) |\psi^- \rangle_{AB}\\
=&\frac{1}{\sqrt{2}} (U|H\rangle_A\otimes|V\rangle_B-U|V\rangle_A\otimes|H\rangle_B)\\
=&\frac{1}{\sqrt{2}} ((cos\theta|H\rangle_A+sin\theta e^{i\phi_1}|V\rangle_A)\otimes|V\rangle_B\\
&-(sin\theta e^{i\phi_2}|H\rangle_A-cos\theta e^{i(\phi_1+\phi_2)}|V\rangle_A)\otimes|H\rangle_B)\\
=&\frac{1}{\sqrt{2}} (|H\rangle_A\otimes(-sin\theta e^{i\phi_2}|H\rangle_B+cos\theta|V\rangle_B)\\
&-|V\rangle_A\otimes(-cos\theta e^{i(\phi_1+\phi_2)}|H\rangle_B-sin\theta e^{i\phi_1}|V\rangle_B))\\
=&\frac{1}{\sqrt{2}} (|H\rangle_A\otimes U'|V\rangle_B-|V\rangle_A\otimes U'|H\rangle_B)\\
=& (I\otimes U') |\psi^- \rangle_{AB},\\
\end{split}
\]
where
\[
U'=\left(
      \begin{array}{cc}
       -cos\theta e^{i(\phi_1+\phi_2)} & -sin\theta e^{i\phi_2}\\
       -sin\theta e^{i\phi_1} & cos\theta\\
      \end{array}
      \right).
\]
and $U'U=-e^{i(\phi_1+\phi_2)}I$. As a result, the effect of $U$ operating on photon A is equivalent to $U^{-1}$ operating on photon B up to a global phase.

Photon A then passes through the beam displacer BD1 and two HWPs fixed at $0^\circ$ and $45^\circ$ in upper path ($u$) and lower path ($l$) respectively. Beam displacer is used to cause the horizontally polarized component to walk off, whereas the vertically polarized component is transmitted unperturbed. 
%
%Photon 1 then passes through the beam displacer BD1, which causes the horizontally polarized component to walk off, whereas the vertically polarized component is transmitted unperturbed. 
%
%As a consequence, the polarization degree of freedom(DOF) and the path DOF of photon 1 get entangled.
%
% After passing the two HWPs fixed at $0^\circ$ and $45^\circ$ in upper path ($u$) and lower path ($l$) respectively, photon 1 achieves the conversion of DOF between polarization and path. 
As a result, the effect of BD1 and the two HWPs is to add a path qubit on photon A, converting $|H\rangle_A$ to $ |H\rangle_A \otimes|u\rangle_A$, and $|V\rangle_A$  to $|H\rangle_A \otimes|l\rangle_A$. After the two HWPs, the quantum state of the two photons can be written as
\[
|H\rangle_A \otimes\frac{1}{\sqrt{2}}(|u\rangle_A\otimes U^{-1}|V\rangle_B-|l\rangle_A\otimes U^{-1}|H\rangle_B),
\]
where the path qubit of photon A and the polarization qubit of photon B are entangled, and the polarization qubit of photon A is separable to the other two qubits.
QWP1 and HWP1 are then used to prepare the polarization qubit of photon A to the generic superposition state $|\varphi \rangle_{A} = \alpha |H\rangle_A + \beta |V\rangle_A$, and the two photon state becomes
\begin{equation}
|\varphi \rangle_{A} \otimes\frac{1}{\sqrt{2}}(|u\rangle_A\otimes U^{-1}|V\rangle_B-|l\rangle_A\otimes U^{-1}|H\rangle_B)
\end{equation}

Comparing with the theoretical scheme, there is a one-to-one correspondence between the polarization qubit of photon A (the path qubit of photon A/ the polarization qubit of photon B) and qubit 1 (qubit 2/ qubit 3) shown in Fig. 1.
%The unknown unitary, realized by a sandwich-like structure composed of two quarter waveplates~(QWPs) and a half waveplate~(HWP) in box, is operated on photon A to obtain $|\psi_U \rangle_{AB} = (U\otimes I) |\psi^- \rangle_{AB}$ which is equivalent to $(I\otimes U^{-1} ) |\psi^- \rangle_{AB}$ up to a global phase, as shown in eq.(3).
%The input qubit, encoded in the polarization of photon 1, can prepared in the generic superposition $|\varphi_{in} \rangle = \alpha |H\rangle + \beta |V\rangle $ by tuning QWP1 and HWP1.

Eq. (3) can be expanded into the following form,
\begin{equation}
\begin{split}
&(\alpha |H\rangle_A + \beta |V\rangle_A) \otimes\frac{1}{\sqrt{2}}(|u\rangle_A\otimes U^{-1}|V\rangle_B-|l\rangle_A\otimes U^{-1}|H\rangle_B)\\
&=\frac{1}{\sqrt{2}}(\alpha |H\rangle_A|u\rangle_A \otimes U^{-1}|V\rangle_B+\beta |V\rangle_A|u\rangle_A\otimes U^{-1}|V\rangle_B)\\
&\qquad -\alpha |H\rangle_A|l\rangle_A\otimes U^{-1}|H\rangle_B-\beta |V\rangle_A|l\rangle_A\otimes U^{-1}|H\rangle_B))\\
&=\frac{1}{2\sqrt{2}}(|H\rangle_A|u\rangle_A+|V\rangle_A|l\rangle_A)\otimes U^{-1}(-\beta|H\rangle_B+\alpha|V\rangle_B)\\
&\qquad +\frac{1}{2\sqrt{2}}(|H\rangle_A|u\rangle_A-|V\rangle_A|l\rangle_A)\otimes U^{-1}(\beta|H\rangle_B+\alpha|V\rangle_B)\\
&\qquad -\frac{1}{2\sqrt{2}}(|H\rangle_A|l\rangle_A+|V\rangle_A|u\rangle_A)\otimes U^{-1}(\alpha|H\rangle_B-\beta|V\rangle_B)\\
&\qquad -\frac{1}{2\sqrt{2}}(|H\rangle_A|l\rangle_A-|V\rangle_A|u\rangle_A)\otimes U^{-1}(\alpha|H\rangle_B+\beta|V\rangle_B)\\
&=\frac{1}{2}|\phi^{+}\rangle_{A} \otimes U^{-1}XZ|\varphi \rangle_{B}+\frac{1}{2}|\phi^{-}\rangle_{A} \otimes U^{-1}X|\varphi \rangle_{B}\\
&\qquad -\frac{1}{2}|\psi^{+}\rangle_{A} \otimes U^{-1}Z|\varphi \rangle_{B}-\frac{1}{2}|\psi^{-}\rangle_{A} \otimes U^{-1}|\varphi \rangle_{B},
\end{split}
\end{equation}
where $|\phi^{\pm}\rangle_{A}$ and $|\psi^{\pm}\rangle_{A}$ are the Bell states of the polarization qubit and path qubit of photon A, and $|\varphi \rangle_{B}=\alpha|H\rangle_B+\beta|V\rangle_B$ is the generic state of photon B.

From eq. (4), it is clear that the desired state $U^{-1}|\varphi \rangle_{B}$ can be obtained by projecting the polarization qubit and the path qubit of photon A into the singlet state $|\psi^{-}\rangle_{A}=\frac{1}{\sqrt{2}}(|H \rangle_{A}|l\rangle_A-|V \rangle_{A}|u\rangle_A)$.
Such state projection is realized by BD2, a HWP at $67.5^\circ$, a polarization beamsplitter~(PBS1) and a single-photon detector D1 as shown in Fig. 2. This can be understood as follows: BD2 can convert $\frac{1}{\sqrt{2}}(|H \rangle_{A}|l\rangle_A-|V \rangle_{A}|u\rangle_A)$ to $\frac{1}{\sqrt{2}}(|H \rangle_{A}-|V \rangle_{A})\otimes|u\rangle_A$; a HWP at $67.5^\circ$ then converts $\frac{1}{\sqrt{2}}(|H \rangle_{A}-|V \rangle_{A})$ to $|H \rangle_{A}$, which can go through PBS1 and be detected by D1. When D1 clicks, the singlet projection succeeds and photon B becomes the desired $U^{-1}|\varphi\rangle_B$, which is detected by a polarization analyzer consisting of QWP2, HWP2, PBS2 and D2. Here we note that, as path qubit is used in the experiment, the setup needs a phase stability at optical wavelength level, which is guaranteed by constructing the interferometer with beam displacers \cite{B03,Z19}.

In our experiments, we tested the scheme using the following three single-qubit unitary operations,
\begin{align}\nonumber
U_1=\left(
      \begin{array}{ccc}
        \frac{1}{2} \ \  & \frac{\sqrt{3}}{2} \\
        -\frac{\sqrt{3}}{2}\ \  & \frac{1}{2}
      \end{array}
    \right)\quad ,\quad
U_2=\left(
      \begin{array}{ccc}
        1\ \  & 0 \\
        0\ \  & e^{i \frac{4\pi}{3}}
      \end{array}
    \right)
\end{align}
\begin{align}\nonumber
U_3=\frac{1}{2}\left(
      \begin{array}{ccc}
        -1-i\ \  & 1+i \\
        -1+i\ \  & -1+i
      \end{array}
    \right) . 
\end{align}
The success of reversing the unitaries above is witnessed by a coincidence detection.

In order to assess the implementation of the inverse unitary, we have applied the scheme to different input preparations and performed process tomography at the output. The selected input states have been $|H\rangle$, $|V\rangle$, $|D\rangle=\frac{1}{\sqrt{2}}(|H\rangle+|V\rangle)$, $|A\rangle=\frac{1}{\sqrt{2}}(|H\rangle-|V\rangle)$, $|R\rangle=\frac{1}{\sqrt{2}}(|H\rangle+i|V\rangle)$, and $|L\rangle=\frac{1}{\sqrt{2}}(|H\rangle-i|V\rangle)$. For each input state, the output state of qubit 2 has been then measured using three mutually unbiased bases $\{|H\rangle, |V\rangle\}$, $\{|D\rangle, |A\rangle\}$ and $\{|R\rangle, |L\rangle\}$. Upon exploiting data from the above measurements, we may perform QPT and fully characterize the inverse unitary operation by reconstructing the corresponding CP-map.

\begin{figure*}[htbp]
  \center
   \includegraphics[scale=0.52]{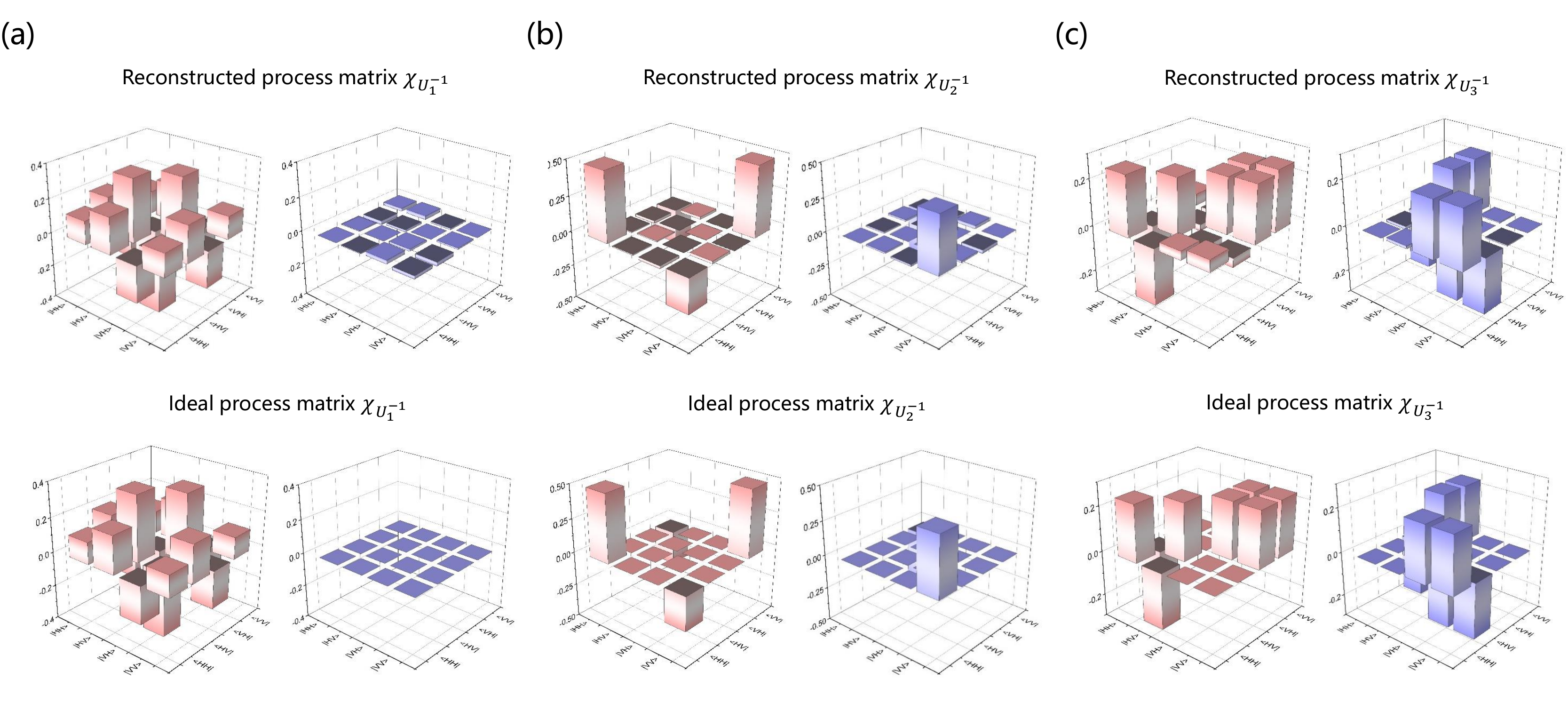}
   \caption{   
   Experimental results: The reconstructed process matrices are shown on the top panels and the ideal process matrices are shown on the bottom panels. The red color and blue color graph represent real and imaginary elements of the reconstructed process matrices $\chi$, respectively. (a) shows the process matrix $\chi_{U^{-1}_1}$. (b) shows the process matrix $\chi_{U^{-1}_2}$. (c) shows the process matrix $\chi_{U^{-1}_3}$.
   }
\end{figure*}

According to Jamiolkowski-Choi isomorphism\cite{J72,C75}, a single qubit quantum operation $\xi$ may be represented by a positive semidefinite operator $\chi$ on the Hilbert spaces of two-qubit state. In particular, a trace preserving map corresponds to a density operator and its density matrix $\chi$ may be obtained from an initial maximally-entangled state $| \phi^+ \rangle=\frac{1}{\sqrt{2}} (|HH\rangle +|VV\rangle)$ by applying the operation $\xi$ on one of the qubits, $i.e.$ $\chi=I\otimes\xi(| \phi^+ \rangle\langle \phi^+ |)$. In turn, the action of the map may be expressed as $\rho_{out}=Tr_{in}[(\rho^T_{in}\otimes I_{out})\chi]$, where $T$ denotes the transposition in a fixed basis. 

We employ maximum likelihood estimation\cite{H97,R05} to reconstruct the quantum process matrix $\chi$ and use fidelity to compare results with the theoretical process matrix representing the unitary operation
\begin{align}
F(\chi)=\frac{Tr[\chi\chi_{ideal}]}{Tr[\chi]Tr[\chi_{ideal}]},
\end{align}
where $\chi_{ideal}$ is the theoretical process matrix representing the unitary operation $V$, i.e.
\begin{align}
\chi_{ideal}=I\otimes V(| \phi^+ \rangle\langle \phi^+ |)I\otimes V^\dagger.
\end{align}

The reconstructed matrices and the ideal matrices corresponding to the inverse unitary operations $U_1^{-1}$, $U_2^{-1}$ and $U_3^{-1}$ are shown in Fig.3.  The corresponding fidelities, calculated using eq.(3), are given by  $F(U^{-1}_1)$=0.9778$\pm$0.0042, $F(U^{-1}_2)$=0.9772$\pm$0.0071, and $F(U^{-1}_3)$=0.9752$\pm$0.0032. The average process fidelity is $F(U^{-1})$=0.9767$\pm$0.0048, which is in good agreement with the theoretical prediction.

\emph{Conclusion-} In summary, we have experimentally demonstrated the inversion of an unknown single-qubit unitary $U$. Our results prove the validity and feasibility of the reversing unitary scheme\cite{QD19}. 
The inverse operation serves as the essential part of HHL algorithm\cite{HHL}, which can be used to solve linear systems of equations and be applied to quantum machine learning \cite{BW17}.
Apart from the scheme proposed by Quintino \emph{et al}\cite{QD19}, there are other universal unitary inversion protocols that may have applications in quantum communications\cite{NC97,BR09}, quantum memory storage\cite{SB19} and Hamiltonian evolution\cite{TD19} in quantum dynamics.

\begin{acknowledgments}
This work was supported by the National Key Research and Development Program (2016YFA0301700 and 2017YFA0305200), the Key Research and Development Program of Guangdong Province of China (2018B030329001 and 2018B030325001), the National Natural Science Foundation of China (Grant No.61974168) and the Natural Science Foundation of Guangdong Province of China (2016A030312012). X.Z. acknowledges support from the National Young 1000 Talents Plan. 
\end{acknowledgments}

%\end{spacing}


\begin{thebibliography}{99}

%


\bibitem{NC00} M. A. Nielsen and I. L. Chuang, Quantum computation and quantum information, Cambridge University Press (2000).
\bibitem{M18} M. Navascues, Resetting uncontrolled quantum systems, Phys. Rev. X 8, 031008 (2018).
\bibitem{Reset2} Z. Li,1,  X. Yin,  Z. Wang, L. Liu, R. Zhang, Y.  Zhang,  X. Jiang, J. Zhang, L. Li,  N. Liu,  X. Zhu, F. Xu, Y. Chen, and J.W. Pan, Photonic realization of quantum resetting, arXiv:1911.11585v1 (2019).
\bibitem{CN97} I. L. Chuang and M. A. Nielsen, Prescription for experimental determination of the dynamics of a quantum black box, J. Mod. Opt. 44, 2455 (1997).
\bibitem{CE06} G. Chiribella and D. Ebler, Optimal quantum networks and one-shot entropies, New J. Phys. 18, 093053 (2016).
\bibitem{SC12} I. S. B. Sardharwalla, T. S. Cubitt, A. W. Harrow, and N. Linden, Universal refocusing of systematic quantum noise, arXiv:1602.07963 (2016).
\bibitem{KSV02} A. Y. Kitaev, A. Shen, and M. N. Vyalyi, Classical and quantum computation, Graduate Studies in Mathematics, American Mathematical Society Providence (2002).
\bibitem{DN05} C. M. Dawson and M. A. Nielsen, The Solovay-Kitaev algorithm, arXiv: 0505030 (2005).
\bibitem{QD19} M. T. Quintino, Q. Dong, A. Shimbo, A. Soeda, and M. Murao, Reversing unknown quantum transformations: A universal protocol for inverting general unitary operations, Phys. Rev. Lett.123, 210502 (2019).
\bibitem{QD019} M. T. Quintino, Q. Dong, A. Shimbo, A. Soeda, and M. Murao, Probabilistic exact universal quantum circuits for transforming unitary operations, Phys. Rev. A 100, 062339 (2019).
\bibitem{BB93} C. H. Bennett, G. Brassard, C. Crépeau, R. Jozsa, A. Peres, and W. K. Wootters, Teleporting an unknown quantum state via dual classical and Einstein-Podolsky-Rosen channels, Phys. Rev. Lett. 70, 1895 (1993).
\bibitem{GC99} D. Gottesman and I. L. Chuang, Demonstrating the viability of universal quantum computation using teleportation and single-qubit operations, Nature 402, 390 (1999).
\bibitem{MS19} J. Miyazaki, A. Soeda, and M. Murao, Complex conjugation supermap of unitary quantum maps and its universal implementation protocol, Phys. Rev. Res. 1, 013007 (2019).
\bibitem{B03} J. L. O’Brien, G. J. Pryde, A. G. White, T. C. Ralph, and  D. Branning, Demonstration of an all-optical quantum controlled-NOT gate, Nature 426, 264 (2003).
\bibitem{Z19}C. Zhang, S. Cheng, L. Li, Q.Y. Liang, B.H. Liu, Y.F. Huang, C.F. Li, G.C. Guo, M.J.W. Hall, H.M. Wiseman, and G.J. Prydeo, Experimental validation of quantum steering ellipsoids and tests of volume monogamy relations, Phys. Rev. Lett. 122, 070402 (2019).
\bibitem{J72} A. Jamiołkowski, Linear transformations which preserve trace and positive semidefiniteness of operators, Reports on Mathematical Physics, 3,275 (1972).
\bibitem{C75} M.-D. Choi, Completely positive linear maps on complex matrices, Linear Algebra and its Applications, 10,285 (1975).
\bibitem{H97} Z. Hradil, Quantum-state estimation, Phys. Rev. A 55, R1561(1997).
\bibitem{R05} R. Blume-Kohout, Hedged maximum mikelihood quantum state estimation, Phys. Rev. Lett. 105, 200504 (2005).
\bibitem{HHL} A. W. Harrow, A. Hassidim, and S. Lloyd, Quantum algorithm for linear systems of equations,  Phys. Rev. Lett. 103, 150502 (2009).
\bibitem{BW17}J. Biamonte, P. Wittek, N. Pancotti, P. Rebentrost, N. Wiebe and S. Lloyd, Quantum machine learning, Nature 549, 195 (2017). 
\bibitem{NC97} M. A. Nielsen and C. M. Caves, Reversible quantum operations and their application to teleportation, Phys. Rev. A 55, 2547 (1997).
\bibitem{BR09} S. D. Bartlett, T. Rudolph, R. W. Spekkens, and P. S. Turner, Quantum communication using a bounded-size quantum reference frame, New J. Phys. 11, 063013 (2009).
\bibitem{SB19} M. Sedlák, A. Bisio, and M. Ziman, Optimal probabilistic storage and retrieval of unitary channels, Phys. Rev. Lett.122.170502 (2019).
\bibitem{TD19} D. Trillo, B. Dive, and M. Navascu\'es, Remote time manipulation, arXiv:1903.10568 (2019).





%\bibitem{NC97} Nielsen M A, Chuang I L. Programmable quantum gate arrays[J]. Physical Review Letters, 1997, 79(2): 321.
%\bibitem{VMC02} Vidal G, Masanes L, Cirac J I. Storing quantum dynamics in quantum states: a stochastic programmable gate[J]. Physical review letters, 2002, 88(4): 047905.
%\bibitem{HBZ02} Hillery M, Bu\v zek V, Ziman M. Probabilistic implementation of universal quantum processors[J]. Physical Review A, 2002, 65(2): 022301.
%\bibitem{OPW03} O'Brien J L, Pryde G J, White A G, et al. Demonstration of an all-optical quantum controlled-NOT gate[J]. Nature, 2003, 426(6964): 264.
%\bibitem{HRZ04} Huang Y F, Ren X F, Zhang Y S, et al. Experimental teleportation of a quantum controlled-NOT gate[J]. Physical review letters, 2004, 93(24): 240501.
%\bibitem{BB93} Charles H.Bennett, Gilles Brassard, et al. Teleporting an Unknown Quantum State via Dual Classical and Einstein-Podolsky-Rosen Channels[J]. Physical review letters, 1993, 70(13): 1895.
%\bibitem{BPK99} Dik Bouwmeester, Jian-Wei Pan, Klaus Mattle, et al. Experimental quantum teleportation[J]. Nature, 1999, 390: 575.
%\bibitem{MJM08} Mičuda M, Je\v zek M, Du\v sek M, et al. Experimental realization of a programmable quantum gate[J]. Physical Review A, 2008, 78(6): 062311.
%\bibitem{SJF09} Slodi\v cka L, Je\v zek M, Fiur\'a\v sek J. Experimental demonstration of a teleportation-based programmable quantum gate[J]. Physical Review A, 2009, 79(5): 050304.
%\bibitem{YFL10} Yao X C, Fiur\'a\v sek J, Lu H, et al. Experimental realization of programmable quantum gate array for directly probing commutation relations of Pauli operators[J]. Physical review letters, 2010, 105(12): 120402. 
%\bibitem{HZB022} Hillery M, Ziman M, Bu\v zek V. Implementation of quantum maps by programmable quantum processors[J]. Physical Review A, 2002, 66(4): 042302.
%\bibitem{HZB04} Hillery M, Ziman M, Bu\v zek V. Improving the performance of probabilistic programmable quantum processors[J]. Physical Review A, 2004, 69(4): 042311.
%\bibitem{HZB06} Hillery M, Ziman M, Bu\v zek V. Approximate programmable quantum processors[J]. Physical Review A, 2006, 73(2): 022345.
%\bibitem{BBK05} Brazier A, Bu\v zek V, Knight P L. Probabilistic programmable quantum processors with multiple copies of program states[J]. Physical Review A, 2005, 71(3): 032306.
%\bibitem{CC05} Lin Chen, Yi-Xin Chen. Probabilistic implementation of a nonlocal operation using a nonmaximally entangled state[J]. Physical Review A, 2005, 71(5): 054302.
%\bibitem{FDF02} Fiur\'a\v sek J, Du\v sek M, Filip R. Universal measurement apparatus controlled by quantum software[J]. Physical review letters, 2002, 89(19): 190401.
%\bibitem{DB02} Du\v sek M, Bu\v zek V. Quantum-controlled measurement device for quantum-state discrimination[J]. Physical Review A, 2002, 66(2): 022112.
%\bibitem{HVC01} Huelga S F, Vaccaro J A, Chefles A, et al. Quantum remote control: teleportation of unitary operations[J]. Physical Review A, 2001, 63(4): 042303.%%
%\bibitem{HRZ04} Y.-F. Huang, X.-F. Ren, Y.-S. Zhang, L.-M. Duan, and G.-C. Guo, Experimental teleportation of a quantum controlled-NOT gate[J].  Phys. Rev. Lett. 93, 240501 (2004).
%\bibitem{MJM08} M. Mičuda, M. Ježek, M. Dušek, and J. Fiurášek, Experimental realization of a programmable quantum gate[J]. Phys. Rev. A 78, 062311 (2008).
%\bibitem{SJF09} L. Slodička, M. Ježek, and J. Fiurášek, Experimental demonstration of a teleportation-based programmable quantum gate[J], Phys. Rev. A 79, 050304 (2009).
%\bibitem{YFL10} X.-C. Yao, J.  Fiur\'a\v sek J, H. Lu, W.-B. Gao, Y.-A. Chen, Z.-B. Chen, and J.-W. Pan, Experimental realization of programmable quantum gate array for directly probing commutation relations of Pauli operators[J], Phys. Rev. Lett. 105.120402 (2010).

%\bibitem{V55} J. von Neumann, Mathematical foundations of quantum mechanics[M], Princeton University Press (1955).
%\bibitem{UK92} M. Ueda and M. Kitagawa, Reversibility in quantum measurement processes[J], Phys. Rev. Lett. 68.3424 (1992).
%\bibitem{I93} A. Imamoglu, Logical reversibility in quantum-nondemolition measurements[J], Phys. Rev. Lett. 47.4577 (1993).
%\bibitem{R94} A. Royer, Reversible quantum measurements on a spin 1/2 and measuring the state of a single system[J], Phys. Rev. Lett. 73.913 (1994).
%\bibitem{KJ06} A. N. Korotkov and A. N. Jordan, Undoing a weak quantum measurement of a solid-state qubit[J], Phys. Rev. Lett. 97.166805 (2006).
%\bibitem{TU07} H. Terashima and M. Ueda, Probabilistic reversing operation with fidelity and purity gain for macroscopic quantum superposition[J], Phys. Rev. A 75, 052323 (2007).
%\bibitem{SA09} Q. Sun, M. Al-Amri, and M. S. Zubairy, Reversing the weak measurement of an arbitrary field with finite photon number[J], Phys. Rev. A 80, 033838 (2009).







\end{thebibliography}
\end{document}